# Functional Brain Network Identification in Opioid Use Disorder Using Machine Learning Analysis of Resting-State fMRI BOLD Signals

Ahmed Temtam[a], Megan A. Witherow[a], Liangsuo Ma[b,c], M. Shibly Sadique[a], F. Gerard Moeller[b,c,d,e,f], and Khan M. Iftekharuddin*[a,g]

[a] Vision Lab, Dept. of Electrical Engineering, Old Dominion University, Norfolk, VA, USA

[b] Institute of Drug and Alcohol Studies, Richmond, VA, USA

[c] Department of Psychiatry, Virginia Commonwealth University, Richmond, VA, USA

[d] Department of Pharmacology and Toxicology, Virginia Commonwealth University, VA, USA

[e] Department of Neurology, Virginia Commonwealth University, VA, USA

[f] C. Kenneth and Dianne Wright Center for Clinical and Translational Research, Virginia Commonwealth University, VA, USA

[g] Data Science Institute, Old Dominion University, Virginia Beach, VA, USA

* Corresponding Author (phone: 757-683-4967, fax: 757-683-3898, email: kiftekha@odu.edu)

**Abstract**

Understanding the neurobiology of opioid use disorder (OUD) using resting-state functional magnetic resonance imaging (rs-fMRI) may help inform treatment strategies to improve patient outcomes. Recent literature suggests time-frequency characteristics of rs-fMRI blood oxygenation level-dependent (BOLD) signals may offer complementary information to traditional analysis techniques. However, existing studies of OUD analyze BOLD signals using measures computed across all time points. This study, for the first time in the literature, employs data-driven machine learning (ML) for time-frequency analysis of local neural activity within key functional networks to differentiate OUD subjects from healthy controls (HC). We obtain time-frequency features based on rs-fMRI BOLD signals from the default mode network (DMN), salience network (SN), and executive control network (ECN) for 31 OUD and 45 HC subjects. Then, we perform 5-fold cross-validation classification (OUD vs. HC) experiments to study the discriminative power of functional network features while taking into consideration significant demographic features. The DMN and SN show the most discriminative power, significantly ($p < 0.05$) outperforming chance baselines with mean F1 scores of 0.7097 and 0.7018, respectively, and mean AUCs of 0.8378 and 0.8755, respectively. Follow-up Boruta ML analysis of selected time-frequency (wavelet) features reveals significant ($p < 0.05$) detail coefficients for all three functional networks, underscoring the need for ML and time-frequency analysis of rs-fMRI BOLD signals in the study of OUD.

**Keywords**—opioid use disorder, resting-state fMRI, machine learning, functional brain networks, default mode network, salience network, executive control network

# 1. Introduction

Opioid use disorder (OUD) is a chronic neurological disorder associated with opioid dependence, relapse, and consequent adverse effects to an individual's physical and psychosocial functioning. According to the U.S. Centers for Disease Control and Prevention, approximately 82,000 drug overdose deaths in 2022 involved opioids (76% of all drug overdose deaths) [1]. The U.S. Joint Economic Committee estimates that economic impacts of the opioid crisis in 2020 alone total to nearly $1.5 trillion, accounting for costs associated with healthcare, public safety, productivity loss, lower quality of life, and loss of lives [2]. Currently, there is no cure for OUD. Treatments such as opioid agonist therapy (e.g., methadone, buprenorphine) are effective in reducing withdrawal and risk of overdose, but face difficulties in initiation and continuation of treatment [3]. Furthermore, opioid agonists do not ameliorate opioid dependence and are associated with increased risk of death due to overdose following treatment termination [3]. A critical need for alternative treatment strategies has led to increased interest in neuroimaging to identify targets for new pharmaceutical, behavioral, or neuromodulation-based therapies [4].

To support the identification of treatment targets, resting-state functional magnetic resonance imaging (rs-fMRI) can provide insight into the functional organization of the brain and how it is disrupted by OUD. However, the majority of rs-fMRI studies of OUD study relationships between networks involved in functions such as decision-making, assigning salience to stimuli, and introspective processes. Few have investigated within-network brain activity to localize anomalous neural signals to specific functional networks [5-10]. Furthermore, rs-fMRI studies investigating OUD have mostly relied on traditional methods for rs-fMRI analysis, which assume functional brain networks are stationary during rest and aggregate information over all time points of the rs-fMRI scan [11]. Recently, studies have shown that non-spurious temporal variations occur in rs-fMRI signals over the course of a scan [11, 12]. Thus, machine learning (ML) approaches have been proposed to learn meaningful patterns in time-frequency analysis of rs-fMRI signals [13-15]. However, such data-driven ML-based approaches have not yet been explored for OUD.

In the following sections, we review traditional methods for rs-fMRI analysis, discuss the application of ML in clinical research, and describe the findings and limitations of existing rs-fMRI studies of OUD. Then, we propose data-driven ML-based analysis of rs-fMRI signals to identify important functional networks and time-frequency features for OUD, with the goal of supporting downstream research into new treatment strategies.

1.1 Methods for Resting-State fMRI Analysis

Rs-fMRI captures fluctuations in the blood oxygen-level dependent (BOLD) signal, measuring spontaneous neural activity across the brain while patients are at rest [11]. Rs-fMRI is often used to study neural activity within and across intrinsic functional brain networks. These networks represent spatially distributed brain regions responsible for specific brain functions (e.g., attention, motor, visual, language). In healthy individuals, brain regions within these functional networks show correlated patterns of BOLD fluctuations even while patients are at rest [11]. In unhealthy individuals, studying the disruption of BOLD fluctuations across these networks can provide insight into the underlying neurobiology [11].

Traditional methods for rs-fMRI analysis can be categorized as functional connectivity (FC) analysis and voxel-based methods [12]. FC analysis examines correlations among time-series from different spatially distributed regions of the brain to discover functional relationships [12]. FC analysis methods can be subdivided into seed-based, decomposition-based, clustering-based, and graph theory-based methods. Seed-based analysis (SBA) generates correlation maps by correlating the time-series BOLD signal from a priori selected seed voxels with the time-series from all other voxels, across all time points of the scan [12]. Since SBA requires manual selection of the seed voxels, the analysis must be repeated for different seed regions to comprehensively explore relationships among the brain's functional systems [12]. Decomposition-based approaches aim to factorize the data matrix of observed signals from all voxels of the brain into a product of simpler matrices [11]. The most popular decomposition-based approach for rs-fMRI analysis is independent component analysis (ICA) [11]. ICA decomposes the data into a linear combination of a user-specified number of statistically independent components and is typically used to determine the FC of the entire brain across all time points [11]. Domain expertise is required to interpret which ICA-derived components represent neural signals and which represent noise [11]. Clustering methods identify groups of voxels based on the similarity of their time-series [11]. Examples of clustering algorithms include k-means clustering and hierarchical clustering [11]. In graph theory-based methods, functional networks are represented as an undirected graph constructed by down sampling voxels into nodes and building an adjacency matrix based on the correlations of time-series signals from all possible pairs of nodes. The resulting graph represents all possible connections among brain regions, also known as a functional connectome [12].

While FC analysis is concerned with relationships between different brain regions, voxel-based methods examine local neural activity within specific regions or functional networks. Voxel-based methods include regional homogeneity (ReHo) and the amplitude of low frequency fluctuations (ALFF) [12]. ReHo uses Kendall's coefficient of concordance to evaluate the synchronization of the time-series for each voxel with its neighboring voxels, yielding one value for each voxel across all time points [12]. In ALFF, the Fourier transform is applied to the entire time-series signal for each voxel to obtain its frequency domain representation, which is used to obtain its power spectrum [12]. The ALFF value of each voxel is then the average of the square root calculated at each frequency of the power spectrum, normalized by the mean of the ALFF values for all voxels [12]. Recently, the wavelet transform [16] has been proposed as an improvement over the Fourier transform in the calculation of ALFF values [17]. The Fourier transform represents the signal frequencies by decomposing the signal into a set of stationary sinusoidal functions, which are ill-suited for encoding transient signal variations [17]. By contrast, the wavelet transform decomposes the signal into a set of scaled and translated basis functions described by a set of approximation and detail coefficients. These approximation and detail coefficients represent low and high frequency components, respectively, at different time scales, enabling fine-grained time-frequency analysis of more complex, non-stationary signals [17]. Wavelet-ALFF summarizes a large set of wavelet coefficients into a single number by summing the absolute values of the wavelet coefficients at all time points for each frequency point and taking the average over the frequency band [17]. However, due to the aggregation of information across all wavelet coefficients in Wavelet-ALFF, the relative contributions of individual approximation and detail coefficients are lost, impacting interpretability. Therefore, we propose data-driven ML to select important wavelet coefficients for more fine-grained time-frequency analysis of local neural activity within functional networks for OUD.

*1.2 Machine Learning in Clinical Research*

ML is suitable for a broad range of applications in clinical research, including identifying patterns in patient data sets [18-23], assisting in biomarker discovery [24, 25], and developing predictive models [19-21, 26] that may contribute to more precise risk assessments and prognoses. In the context of OUD, ML has been applied for drug repurposing to support treatment development [27, 28] and classification tasks to identify individuals with or at risk for opioid dependence and associated complications based on records data [22, 23, 29, 30]. In [27] and [28], the Gradient Boosting Trees ML algorithm is paired with molecular fingerprints to predict the treatment characteristics of DrugBank compounds on opioid receptors. In [23] and [30], the Random Forest ML algorithm is used to predict opioid substance dependence based on electronic health records and National Survey on Drug Use and Health data, respectively. Commercial claims data has been used to predict patient overdose status using the Gradient Boosting Trees algorithm in [22] and risk of buprenorphine treatment discontinuation using multiple ML algorithms, including Random Forest, in [29]. To our knowledge, ML for time-frequency analysis of rs-fMRI BOLD signals in OUD has yet to be explored.

When working with patient populations, assumptions on the data distributions such as normality, homogeneity of variances, and independence of samples often do not hold. Furthermore, sample sizes are often much smaller compared to those seen in other ML applications. Tree-based algorithms represent a class of non-parametric approaches that are well-suited for clinical data analysis as they are distribution-free, interpretable, and can learn complex, non-linear relationships from data sets of modest size without assuming the independence of samples. Trees are often organized in ensembles where a prediction is made based on the majority vote of multiple trees. Random Forest and Gradient Boosting Trees ML algorithms used in previous studies of OUD are examples of ensembles of decision trees. An extension of Random Forest, the Boruta ML algorithm [31], has proven to be a powerful tool for determining the statistical significance of features used in clinical research (e.g., biomarker discovery [24, 25], gene expression [32, 33], etc.). Recently, the Boruta algorithm has been used to analyze rs-fMRI FC patterns of individuals diagnosed with autism [34] and attention-deficit/hyperactivity disorder (ADHD) [35, 36]. In addition to the significance of individual features, ML classification studies may provide further insight into network-level importance based on the discriminative power of functional network features in distinguishing between patients with a particular diagnosis and HC [34, 37]. Therefore, we anticipate ML-based measures of functional network discriminative power and time-frequency feature significance to provide useful insights into differentiated functional networks in OUD. *1.3 Resting-State fMRI Studies of Opioid Use Disorder*

Recent rs-fMRI studies of substance use disorders focus on the organization of three core neurocognitive networks in the human brain, i.e., the default mode network (DMN), salience network (SN), and executive control network (ECN), and how abnormal function among these three networks underlies substance use disorders [38, 39]. For example, in substance use disorders, processing may be biased towards the DMN rather than the ECN as a result of the SN assigning greater salience to substance-induced rewards [40].

For OUD, the majority of studies examine FC rather than local activity. Woisard et al. [41] use ICA to study the resting-state FC of voxels within the DMN, SN, and ECN for OUD (n=25) and HC (n=25) patients. No significant differences in FC between OUD and HC groups is found for DMN, SN, or right ECN [41]. A cluster of voxels within the left dorsolateral prefrontal cortex of the left ECN is reported to have significantly weaker FC in OUD compared to HC [41]. Using SBA, Abdulaev et al. [42] investigate the resting-state FC between DMN, SN, and ECN for 12 subjects intoxicated with heroin (a type of opioid) and 16 HC. They report decreased FC between the SN and ECN, decreased FC between SN and DMN, and increased FC between the DMN and ECN [42]. Chen et al. [43] study the effects of methadone treatment on FC and withdrawal symptoms in heroin use disorder (HUD), which is a type of OUD. ICA is used to analyze rs-fMRI scans of 37 HUD patients at baseline and after a year of methadone treatment [43]. Findings include increased FC between the medial prefrontal cortex (mPFC) and the left middle temporal gyrus of the DMN, which is associated with reduced withdrawal symptoms [43]. Increased connectivity of the DMN's mPFC with the SN's anterior insular and middle frontal gyrus is associated with increased withdrawal symptoms, which may be related to increased salience of drug-related cues [43]. Wang et al. [44] study interhemispheric FC of the insula (part of the SN) among HUD patients (n=30) undergoing methadone treatment and HC (n=30). They report lower interhemispheric FC with the insula as the seed region for HUD patients compared to HC [44]. Among HUD patients, less deviation in interhemispheric FC is associated with decreased risk of a positive heroin urine test [44]. Dandurand et al. [45] examine resting-state FC in 11 OUD patients before and three days after starting buprenorphine treatment. They apply SBA using three different seed regions, the posterior cingulate cortex (PCC) within the DMN, the anterior insula within the SN, and dorsolateral prefrontal cortex within the ECN [45]. Greater FC of the DMN and SN is found after initiation of treatment compared to the baseline scan [45]. Mill et al. [46] explore univariate and multivariate approaches based on whole brain resting-state FC and structural MRI to classify prescription OUD (POUD) patients (n=26) and HC (n=21). Using a minimum-distance classification approach with Pearson correlation FC and structural MRI features, they achieve 66.7% accuracy in distinguishing between POUD and HC [46].

Few rs-fMRI studies have studied local neural activity within brain networks for OUD. Such studies have assumed the stationary model of the resting-state BOLD signal and use ReHo or Fourier-based ALFF to measure neural activity in OUD. Studying ReHo of 40 HUD patients before and after one year of methadone treatment, Chang et al. [5] find altered ReHo values in the right caudate within the ECN to be correlated with craving and relapse. Xue et al. [10] report increased

ReHo values in the right orbitofrontal cortex and bilateral posterior central cortex within the ECN for 25 HUD patients under methadone treatment compared to 26 HUD patients after short-term abstinence. In follow-up SBA, they find no significant difference in FC between groups [10]. Liu et al. [7] study differences in ALFF among HC (n=21) and HUD patients (n=21). Comparison of ALFF values for HUD compared to HC reveals increased ALFF in the left middle frontal gyrus of the ECN and decreased ALFF in the left postcentral gyrus, which is considered to have connections to the SN [7]. Wang et al. [9] find decreased ALFF in the right dorsal anterior cingulate cortex (dACC) of the SN, right caudate of the ECN, and right superior medial frontal cortex of heroin-dependent individuals (n=17) compared to HC (n=15) [9]. They report increased ALFF in the cerebellum, left superior temporal gyrus, and left superior occipital gyrus [9]. These six regions with differentiated ALFF are then considered as seeds for FC SBA, finding reduced FC between the right caudate and dorsolateral prefrontal cortex (PFC) associated with the DMN, increased FC between the right caudate and the cerebellum, and abnormal FC between the lateral PFC and dACC in the SN [9]. Compared to HC (n=24), Jiang et al. [6] report heroin-dependent individuals (n=24) show decreased ALFF in the bilateral dACC of the SN, left dorsal lateral prefrontal cortex of the SN, bilateral medial orbit frontal cortex, left middle temporal gyrus, left inferior temporal gyrus, posterior cingulate cortex and left cuneus. They also find increased ALFF in the bilateral angular gyrus, bilateral precuneus, bilateral supramarginal gyrus, left post cingulate cortex, and left middle frontal gyrus [6]. Qiu et al [8] conduct an rs-fMRI study with 14 HC and 14 individuals dependent on cough syrups containing codeine (a type of opioid) before and after receiving 2 weeks of naltrexone treatment. They report that decreased ALFF values in the left post-central gyrus (associated with the SN) and left middle occipital cortex are associated with reduced severity of codeine withdrawal symptoms [8].

In summary, rs-fMRI studies of OUD and HC have found alterations in FC between DMN, SN, ECN, and other parts of the brain. Differentiated local neural activity based on ReHo and ALFF, which assume a stationary representation of the neural signal at resting-state, has been reported for regions within the ECN and SN. We are not aware of any studies that consider time-frequency analysis of local neural activity for OUD. We expect that ML methods, which excel at learning complex patterns and nonlinear relationships, will prove promising for time-frequency neural signal analysis for OUD.

*1.4 Contributions*

This study is the first in the literature to introduce data-driven ML and rs-fMRI BOLD time-frequency signal analysis to study differentiated local neural activity for OUD within functional brain networks including DMN, SN, and ECN. Most rs-fMRI studies of OUD examine FC or local neural activity using ReHo or ALFF, which assume a stationary model of the resting-state BOLD signal. To our knowledge, the non-stationary characteristics of the BOLD signal have not yet been examined for OUD. Our specific contributions are as follows:

- Our time-frequency features enable fine-grained analysis of the rs-fMRI BOLD signals to localize aberrant neural activity to specific functional networks.
- Our proposed ML-based modeling based on Random Forest learns relevant patterns among the time-frequency features from different functional networks to distinguish OUD and HC.

- We propose multiple ML methods, including Random Forest for OUD vs. HC classification to study functional network discriminative power and the Boruta algorithm to study time-frequency feature significance, and statistical analyses to provide insight into differentiated functional networks in OUD.

## 2. Methods

*2.1 Data set*

All study procedures have been approved by the Virginia Commonwealth University Institutional Review Board under IRB number HM20023630 and have been performed in accordance with the Code of Ethics of the World Medical Association [47]. OUD subjects were recruited from the Richmond, Virginia community outpatient setting by using flyers and in-person recruitment at addiction treatment clinics. To obtain a more representative sample of the OUD population that could be seen clinically, no restrictions regarding current drug use were imposed during recruitment. HC subjects were recruited by using flyers and other advertisements. Written informed consent and a thorough screening were obtained, including a medical history, physical examination, and psychiatric and substance use histories conducted using the Mini International Neuropsychiatric Interview (MINI) for Diagnostic and Statistical Manual version 5 (DSM-5). Inclusion criteria were DSM-5 diagnosis of OUD (for OUD group only) and age between 18 and 70 years. Exclusion criteria were any history of schizophrenia, seizure disorder, significant head trauma, any changes to psychoactive medications within 30 days prior to the study period, any other DSM-5 Substance Use Disorder with a severity diagnosis greater than the subject's DSM-5 OUD severity diagnosis (Mild, Moderate, Severe), or DSM-5 Severe Alcohol Use Disorder. In addition, HC subject exclusion criteria were any history of substance use disorder. Subjects were seen for a screening visit and three additional visits in which they completed several behavioral measures, an MRI screening and mock MRI session, and an MRI scan. Participants were asked to refrain from smoking 1 hour before and drinking caffeine 3 hours before their MRI scan. Urine drug screens (UDS) and alcohol breath screens were obtained during each visit. 31 OUD and 45 HC subjects were included in the final analysis. Table 1 provides the diagnostic (OUD or HC) breakdown of the subjects by gender and the mean age for each group. The OUD group (40±11 years) is older than the HC group (33±14 years) (Student t=2.3, df=74, p=0.0225). Table 2 describes the demographic features collected for these subjects.

**Table 1**
Data set Distribution

| Subjects | Male | Female | Total | Mean Age |
|---|---|---|---|---|
| HC | 21 | 24 | 45 | 33.8 |
| OUD | 18 | 13 | 31 | 39.6 |
| Total | 39 | 37 | 76 | |

**Table 2**
Demographic Features

| Patient Demographics | |
|---|---|
| **ETHNICITY** | Ethnicity of the subject |
| **EDUCATION (years)** | Subject's number of years of education |
| **SEX** | Sex of the subject |
| **AGE (years)** | Subject's age at the time of the scan |
| **HEIGHT (inches)** | Height of the subject in inches |
| **WEIGHT (lbs)** | Weight of the subject in pounds |
| **HANDEDNESS** | Dominant hand of the subject |

*2.2 Preprocessing and Feature Extraction*

As described in [41], the rs-fMRI data undergo preprocessing for signal outlier removal, heart rate and respiratory physiologic extraneous signal correction, slice timing correction, spatial smoothing, and anatomical T1 scan registration. Scans that do not meet Parkes et al. [48] quality control criteria for head motion have been removed. FSL MCFLIRT is used to perform head-motion re-alignment and ICA-AROMA is used to perform signal correction [49-51]. The BOLD contrast mechanism captures both neural activity (our target quantity) and the hemodynamic response, the brain's physiological reaction to neural activation [52]. To isolate the neural response from the hemodynamic response, we estimate the resting-state hemodynamic response function (rsHRF) using the rsHRF toolbox by Wu et al. [52]. As the rsHRF represents the sequence of physiological changes that occur in the brain following neural activation, the deconvolution of the rsHRF with the observed BOLD signal yields the predicted underlying neural signal [52]. We apply anatomical masks for DMN, SN, and ECN to isolate neural signals for voxels within each network and average over the spatial dimensions to obtain the representative neural signal for each functional brain network. Using the PyWavelets library [53], we take the discrete wavelet transform (DWT) of the representative neural signals from each functional brain network to obtain wavelet features for time-frequency analysis. The wavelet features consist of approximation and detail coefficients representing different neural frequencies at different time dilations, which are extracted through multi-resolution decomposition using the Daubechies wavelet with four vanishing moments (db4) and three approximation levels (L=3).

*2.3 ML-Based Feature Selection and OUD vs. HC Classification*

We study the discriminative power of the time-frequency features from each of the functional brain networks by performing multiple ML classification experiments. We consider time-frequency features extracted from each of the three functional networks (DMN, SN, ECN) as separate feature sets. For each feature set, we use the Least Absolute Shrinkage and Selection Operator (LASSO) approach to select the most important features based on L1 regularization. The selected features are used to perform classification between OUD and HC.

For classification, we choose the Random Forest ML algorithm, which does not require assumptions regarding independence and normality that do not hold for our data set. The Random Forest algorithm is built on decision trees. A decision tree is a distribution-free, non-parametric model that performs classification by partitioning the data based on decision rules on individual features. These rules are learned based on minimizing errors on a training set. Random Forest models are ensembles of multiple decision trees trained on different randomly sampled subsets of data and features. Each tree makes its own class prediction, and the overall prediction of the model is the majority vote of the trees' individual predictions.

We use Scikit-learn library (https://scikit-learn.org/) implementations of LASSO and Random Forest for our experiments.

*2.4 Boruta ML Algorithm for Feature Significance*

The Boruta algorithm is an ML approach for 'all-relevant' feature selection originally developed for research in genetics where features may be correlated [31], which is also to be expected for our rs-fMRI BOLD time-frequency features. The Boruta algorithm builds a statistical framework around the Random Forest ML algorithm to assess feature importance [31]. For a specified Type I error rate $\alpha$, all significantly discriminative features, e.g., for distinguishing between patients with OUD and HC, are identified based on statistical tests [31]. The Boruta algorithm produces highly stable feature selections and does not require samples to be independent or normally distributed [31, 54].

We use the Boruta ML algorithm to study the statistical significance of the demographic and time-frequency features in distinguishing between OUD and HC. Building upon [31], our specific approach is as follows. First, we build our information system by joining the $p$ features on the index of the $n = 76$ subjects to form a data matrix with shape $n$ samples × $p$ features. Each of the $n = 76$ samples is associated with a label, either 'OUD' or 'HC', which is encoded in a separate label vector. Next, we make $m$ copies of the data matrix. In these copies, we permute the sample values within each feature column to yield randomized 'shadow features', which will be used as references for assessing the significance of the original, unpermuted features. We join the $n \times p$ data matrix and its $m$ permuted replicates to form the extended information system with shape $n \times mp$. Then, the following steps are repeated until the significance of each of the features is determined:

1. Run the Random Forest algorithm on the extended information system to obtain feature importances for the features and shadow features with respect to their relative contributions in minimizing the OUD vs. HC classification loss.
2. Compare the importance of each feature to the maximum of the shadow features' importance. Features with a higher importance than the maximum of the shadow features' importance are assigned a 'hit'.
3. Assign p-values to each of the features using the Binomial distribution at the end of each iteration (e.g., with the null hypothesis $H_0: \eta = 0.5$, where $\eta$ is the probability of a hit, a feature may be assigned $h$ hits in $t$ Bernoulli trials (iterations)). We perform one-tailed binomial tests to confirm ($H_0: \eta > 0.5$) important/significant features and reject ($H_0: \eta < 0.5$) unimportant features. P-values are Bonferroni-corrected to account for multiple testing. Confirmed and rejected features are removed from the information system before continuing with the next iteration.

To conduct our analysis, we use the BorutaPy library (https://github.com/scikit-learn-contrib/boruta_py/) with dynamic selection of the value of $m$ and the number of trees in the ensemble at each iteration based on the number of features in the information system. We choose $α = 0.05$. From the Boruta algorithm, we obtain significant or not significant designations for each of features.

*2.5 Brain Mapping and Visualization*

We apply group analysis to systematically identify and characterize consistent patterns of neural activity across individuals within the context of OUD. The group analysis is performed using SPM (https://www.fil.ion.ucl.ac.uk/spm/doc/) and utilizes automated anatomical labeling atlas 3 (AAL3) [55] to localize clusters of significant voxels to anatomically defined brain regions. This enables us to pinpoint and visualize specific brain regions with differentiated neural activity for OUD subjects.

*2.6 Experimental Pipeline*

Figure 1 summarizes the proposed experimental pipeline. To account for potential biases (e.g., age, sex, education), we first determine the demographic features from Table 2 that are significantly different between the two groups. Due to non-normality of the demographic feature distributions, we use a non-parametric approach based on trees (Boruta) to determine the statistical significance. We preprocess the rs-fMRI scans and extract time-frequency (wavelet) features for time-frequency analysis of the BOLD signals from three functional networks (DMN, ECN, and SN). Next, we split the data into training and testing sets via 5-fold cross validation. The folds are assembled based on stratified sampling such that each fold reflects the class distribution of the overall data set. From the training set, we randomly sample 20% of the samples to use as a validation set. Based on the training set, we use LASSO to perform feature selection on the time-frequency features extracted from each network. We train Random Forest models on the selected features and use the held-out validation set to tune the number of trees in the Random Forest ensemble. We enter the significant group level demographic variables with the selected time-frequency features into the Random Forest models. For comparison, we also train Random Forest models on the time-frequency features only, without accounting for significant demographic features. We perform OUD vs. HC classification studies with the selected time-frequency features for each network to study their discriminative power in distinguishing between OUD and HC test set samples. We perform one-sample t-tests to determine whether the results are statistically significant compared to a baseline random chance classifier. We also perform paired t-tests to investigate the statistical significance of differences in discriminative power among the three functional networks. Using Boruta, we investigate the statistical significance of the time-frequency features selected by the LASSO model in distinguishing between OUD and HC. We visualize the neural activity within each functional network using SPM.

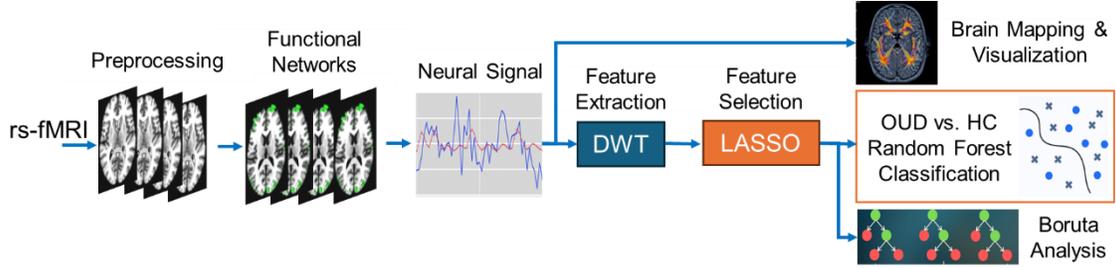

**Fig. 1.** Experimental pipeline for ML-based time-frequency analysis of local neural activity within DMN, SN, and ECN functional network for OUD.

## 3. Results

*3.1 Identification of Significant Demographic Features with Boruta ML Algorithm*

Our Boruta analysis of the demographic features (Table 2) reveals significant ($p < \alpha = 0.05$) differences in age and education between the OUD and HC groups. No significant differences are identified for ethnicity, sex, height, weight, or handedness.

*3.2 ML-based Discriminative Power of Time-Frequency Features from DMN, SN, and ECN*

Based on the validation performance, we set the number of trees in the Random Forest classifier to 100. The 5-fold cross validation test results for each of the three functional networks (DMN, SN, and ECN), considering time-frequency features only (not accounting for age and education) are reported in Table 3. To assess whether any of the functional networks are significantly more discriminative for OUD compared to the others, we perform paired t-tests for each performance metric and pair of functional networks (DMN vs. SN, DMN vs. ECN, SN vs. ECN) using the five-fold test results. The p-values are reported in Table 4. Prior to entering age and education into the model, DMN features yield the best performance in OUD vs. HC classification across all metrics with a mean F1 score of 0.6790, mean AUC of 0.8304, and mean accuracy of 0.7225. The next best performing feature set is SN with a mean F1 score of 0.6053, mean AUC of 0.7733, and mean accuracy of 0.6967. Relative to the DMN and SN feature sets, ECN features yield the lowest classification performance with a mean F1 score of 0.5625, mean AUC of 0.6939, and mean accuracy of 0.6567. Considering a Type I error rate of $\alpha = 0.05$, all feature sets significantly outperform random chance classification across all metrics with the exception of F1 score for ECN. Considering pairwise comparisons of functional network discriminative power, only DMN vs. ECN AUC is significant with a p-value of 0.0470. There are no significant differences in F1 score, AUC, or accuracy for DMN vs. SN or SN vs. ECN, and no significant differences in F1 score or accuracy for DMN vs. ECN.

**Table 3**

5-Fold Cross-Validation Accuracy, Area Under the ROC Curve (AUC), and F1 Scores for Models Trained on DMN, SN, and ECN Features with 95% Confidence Intervals; '*' p<0.05, '**' p<0.01. '***' p<0.001 One-Sample t-test against Chance Classification

| Metric | Network | Mean | 95% Confidence Interval | |
|---|---|---|---|---|
| | | | Lower Bound | Upper Bound |
| Accuracy | DMN | 0.7225* | 0.6010 | 0.8440 |
| | SN | 0.6967** | 0.6339 | 0.7595 |
| | ECN | 0.6567* | 0.5657 | 0.7477 |
| AUC | DMN | 0.8304** | 0.7044 | 0.9564 |
| | SN | 0.7733*** | 0.7203 | 0.8263 |
| | ECN | 0.6939** | 0.6261 | 0.7617 |
| F1 | DMN | 0.6790* | 0.5289 | 0.8291 |
| | SN | 0.6053* | 0.5182 | 0.6924 |
| | ECN | 0.5625 | 0.4322 | 0.6928 |

**Table 4**

Statistical Significance (p-values) of Testing Metrics for DMN, SN, and ECN Features based on Paired t-tests

| Comparison | Accuracy | AUC | F1 |
|---|---|---|---|
| DMN vs SN | 0.5970 | 0.3751 | 0.1997 |
| DMN vs ECN | 0.2351 | 0.0470 | 0.1388 |
| SN vs ECN | 0.3739 | 0.0571 | 0.4379 |

Table 5 reports our evaluation of the relative discriminative power of time-frequency features from each of the three functional networks while also entering significant demographic features (age and education) into the Random Forest models. The p-values of paired t-tests for each performance metric and pair of functional networks (DMN vs. SN, DMN vs. ECN, SN vs. ECN) using the five-fold test results are reported in Table 6. Compared to Table 3, Table 5 shows that the inclusion of age and gender improves mean OUD vs. HC classification performance across all metrics for all three functional networks (DMN, SN, ECN). The best classification performance is obtained using features from the DMN (F1 score 0.7097, AUC 0.8378, accuracy 0.7475) and SN (F1 score 0.7018, AUC 0.8755, accuracy 0.7892), followed by the ECN (F1 score 0.5639, AUC 0.7884, accuracy 0.6975). Considering a Type I error rate of $\alpha = 0.05$, all feature sets significantly outperform random chance classification across all metrics with the exception of accuracy and F1 score for ECN. There is no significant difference in any of the performance metrics between the three networks (DMN vs. SN, DMN vs. ECN, and SN vs. ECN).

**Table 5**
5-Fold Cross-validation Accuracy, Area under the ROC Curve (AUC), and F1 Scores for Models Trained on DMN, SN, and ECN Features with Significant Demographic Features (Age, Education) with 95% Confidence Intervals; '*' p<0.05, '**' p<0.01. '***' p<0.001 One-Sample t-test against Chance Classification

| Metric | Network | Mean | 95% Confidence Interval | |
|---|---|---|---|---|
| | | | Lower Bound | Upper Bound |
| Accuracy | DMN | 0.7475* | 0.6016 | 0.8934 |
| | SN | 0.7892** | 0.7016 | 0.8768 |
| | ECN | 0.6975 | 0.4947 | 0.9003 |
| AUC | DMN | 0.8378** | 0.7045 | 0.9711 |
| | SN | 0.8755** | 0.7599 | 0.9911 |
| | ECN | 0.7884* | 0.6234 | 0.9534 |
| F1 | DMN | 0.7097* | 0.5365 | 0.8829 |
| | SN | 0.7018* | 0.5571 | 0.8465 |
| | ECN | 0.5639 | 0.2024 | 0.9254 |

**Table 6**
Statistical Significance (p-values) of Performance Metrics for DMN, SN, and ECN Features with Significant Demographic Features (Age, Education) based on Paired t-tests

| Comparison | Accuracy | AUC | F1 |
|---|---|---|---|
| DMN vs SN | 0.3936 | 0.3764 | 0.8644 |
| DMN vs ECN | 0.5687 | 0.1439 | 0.3259 |
| SN vs ECN | 0.2696 | 0.1606 | 0.3693 |

*3.3 Significance of ML-selected Time-Frequency Features in OUD vs. HC*

We perform follow-up Boruta analysis to determine the statistical significance of selected time-frequency (wavelet) features from each network in distinguishing between OUD and HC. Considering a Type 1 error rate of $\alpha = 0.05$, the following features are significant: for DMN, level 2 detail coefficient 31; for SN, level 1 detail coefficient 19, level 1 detail coefficient 109, and level 1 detail coefficient 201; and for ECN, level 1 detail coefficient 141, level 1 detail coefficient 144, and approximation coefficient 14.

*3.4 Visualization of OUD-related Neural Activity for DMN, SN, and ECN*

Visualizations based on group analysis are shown in Fig. 2 through Fig. 5. Fig. 2 shows the brain activity for OUD in green over standard anatomical images. Fig. 3, Fig. 4, and Fig. 5 overlay OUD-related neural activity on the DMN, SN, and ECN, respectively.

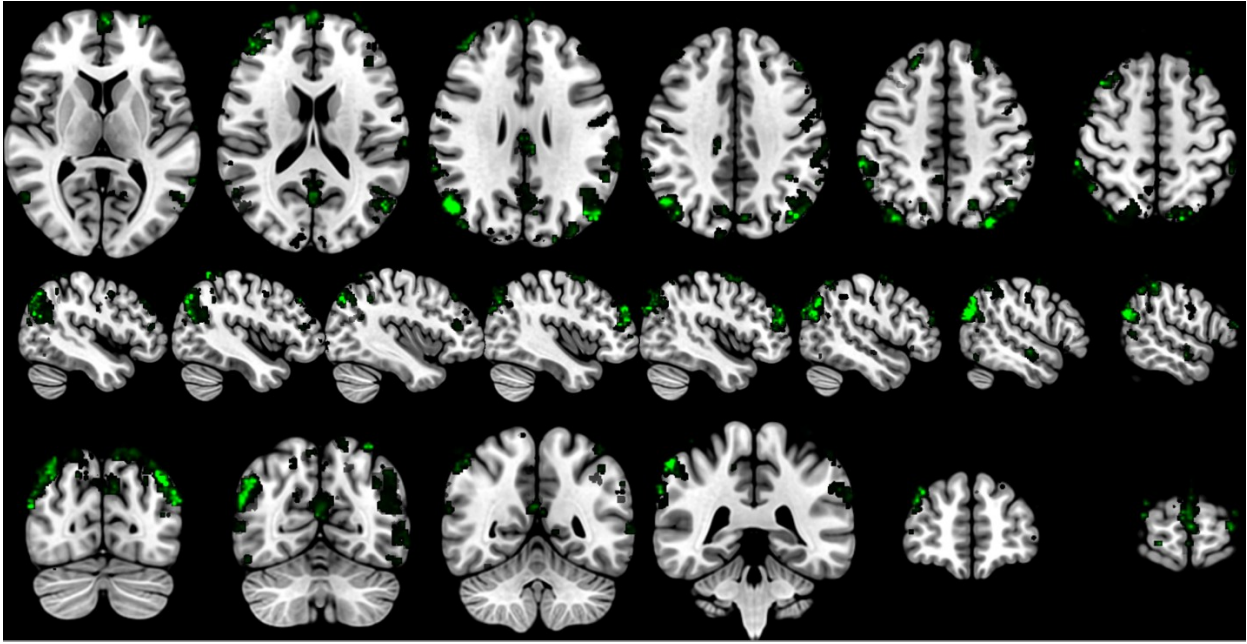

**Fig. 2.** Group analysis showing brain activity (green) for OUD.

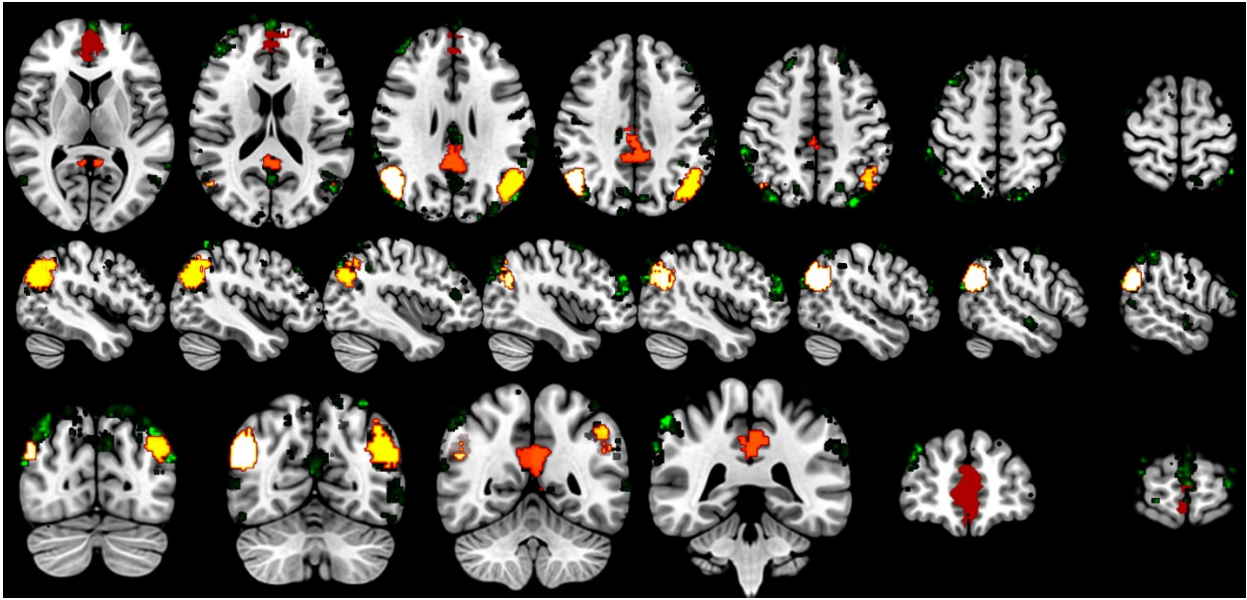

**Fig. 3.** Group analysis illustrating brain activity in OUD (green) overlaid on the DMN (red).

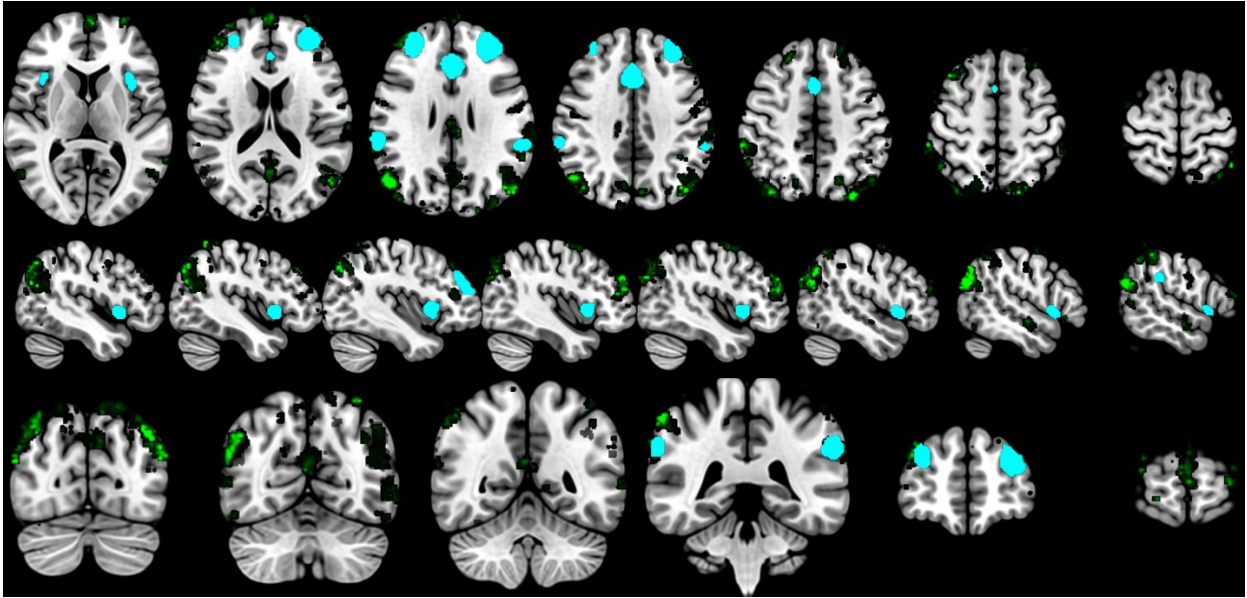

**Fig. 4.** Group analysis illustrating brain activity in OUD (green) overlaid on the SN (light blue).

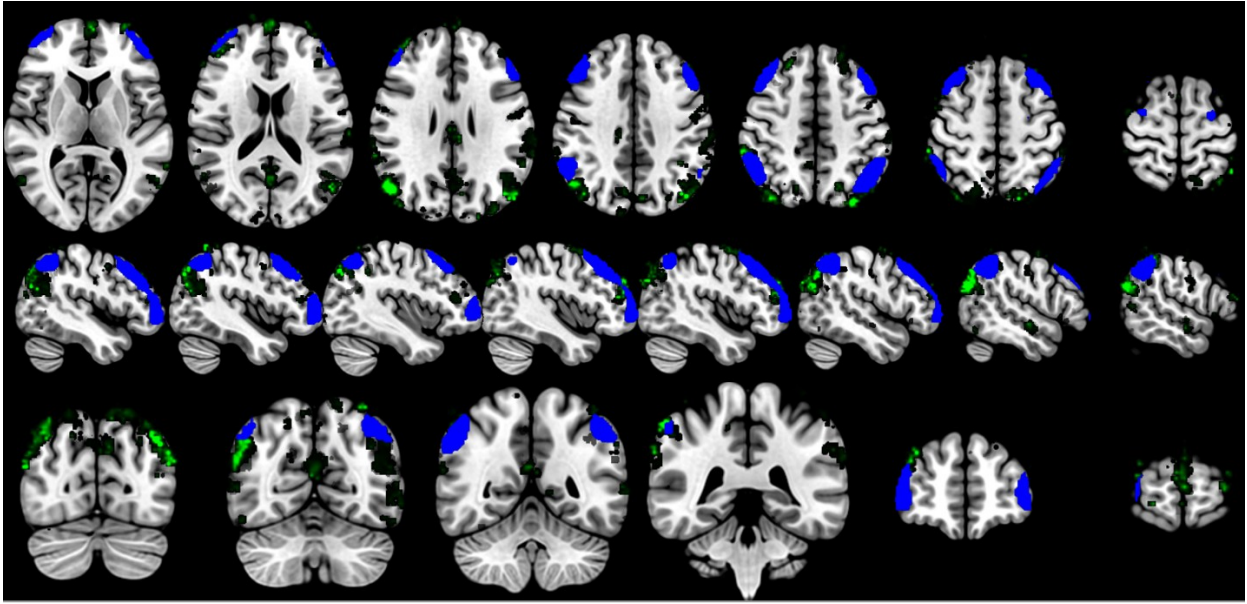

**Fig. 5.** Group analysis illustrating brain activity in OUD (green) overlaid on the ECN (blue).

## 4. Discussion

Advancing our understanding of the neurobiology underlying OUD is a crucial step towards the development of new treatment strategies to improve patient outcomes. Table 7 summarizes existing rs-fMRI studies of OUD in the literature. The majority of rs-fMRI studies of OUD perform FC analyses between different areas of the brain. Few studies investigate local neural activity within functional networks and such studies use measures that assume a stationary representation of the neural signal at resting-state (e.g., ALFF, ReHo). However, neural activity varies meaningfully over the course of an rs-fMRI scan [11, 12]. Our study fills an important gap in the time-frequency analysis of local neural activity within functional brain networks. Our use of ML enables the processing of a large quantity of wavelet coefficients that would typically be averaged into a single value, even using state-of-the-art measures like Wavelet-ALFF [17]. Therefore, our ML-based approach enables more fine-grained analysis of the time-frequency components of the signal.

Our OUD vs. HC Random Forest classification experiments consider time-frequency features extracted from each functional network with significant demographic variables (age and education). Based on our results (Table 5), DMN and SN show the most discriminative power with mean F1 scores of 0.7097 ($p < 0.05$) and 0.7018 ($p < 0.05$), respectively, and mean AUCs of 0.8378 ($p < 0.01$) and 0.8755 ($p < 0.01$), respectively. For ECN, the mean F1 score of 0.5639 is not significantly better than chance ($p > 0.05$), but the mean AUC of 0.7884 is significant ($p < 0.05$). These results suggest that time-frequency analysis of all three functional networks, but especially DMN and SN, may prove useful for the study of OUD. As reported in Table 6, we find no significant differences ($p > 0.05$) in the discriminative power of time-frequency features extracted from each of the three networks (DMN vs. SN, DMN vs. ECN, SN vs. ECN), which suggests that features from all networks may be meaningful for future study.

Our Boruta analysis to identify time-frequency (wavelet) features that are significant ($p < 0.05$) in discriminating between OUD and HC reveals one significant detail coefficient for the DMN, three significant detail coefficients for the SN, and for the ECN, two significant detail coefficients and one significant approximation coefficient. Approximation coefficients represent low-frequency components associated with long-term signal trends, while detail coefficients represent high frequency components such as transient signal alternations. Identification of significant detail coefficients for all three functional networks indicates the presence of meaningful transient signal variations, motivating the utility of time-frequency analysis for local neural signals. Previous studies [5-10] utilizing ReHo and ALFF to study local neural signals in OUD have reported differentiated activity for regions within the ECN and SN. Our time-frequency analysis suggests that the DMN also shows differentiated activity when accounting for the non-stationary nature of the resting-state neural signal.

Further studies are required to better understand the unique and complementary contributions of FC analysis and local neural signal analysis, as well as the implications of findings for patient prognosis. For example, preliminary results by Woisaird et al. [41] suggest that reduced FC in the left ECN may be relevant for characterizing the neurobiology of OUD. Furthermore, future research is needed to investigate the extent to which the DMN, SN, and ECN and their constituent anatomical parts are implicated in OUD. Table 7 further shows that our study benefits from a greater sample size for both OUD (31) and HC (45) compared to similar prior works.

**Table 7**
Rs-fMRI Studies of OUD

| Study | Method | #OUD | #HC | Findings for OUD |
|---|---|---|---|---|
| **Wang et al., 2010 [56]** | FC: SBA | 15 | 15 | Decreased connectivity of the ACC of the SN to the parahippocampal gyrus and PCC of the DMN |
| **Jiang et al., 2011 [6]** | Local: ALFF | 24 | 24 | Decreased ALFF in the bilateral dACC of the SN, left dorsal lateral prefrontal cortex of the SN, and other regions; increased ALFF in bilateral angular gyrus, bilateral precuneus, bilateral supramarginal gyrus, left post cingulate cortex, left middle frontal gyrus |
| **Ma et al., 2011 [57]** | FC: ICA | 14 | 13 | Increased FC in the right hippocampus and decreased FC in right dorsal anterior cingulate cortex and left caudate in the DMN |
| **Liu et al., 2011 [58]** | FC: Graph Theory Analysis | 16 | 16 | Increased FC in the medial frontal gyrus and decreased connectivity in the anterior cingulate cortex of the SN |
| **Li et al., 2013 [59]** | FC: SBA | 14 | 15 | Altered FC in PCC-insula and PCC-striatum areas of the DMN may be regarded as biomarkers of brain damage severity |
| **Wang et al., 2013 [9]** | Local: ALFF | 15 | 15 | Decreased ALFF in right dACC of the SN, right caudate of the ECN, right superior medial frontal cortex; increased ALFF in the cerebellum, left superior temporal and superior occipital gyrus |
| **Jiang et al., 2013 [60]** | FC: Graph Theory Analysis | 15 | 15 | Decreased FC in the ECN, including the left middle frontal gyrus and right precuneus, and increased FC in the left hippocampus |
| **Zhang et al., 2015 [61]** | FC: SBA | 21 | 15 | FC of three subregions of the ACC in the SN are studied; FC variations are found for all three subregions. |
| **Ma et al., 2015 [62]** | FC: SBA | 14 | 14 | FC within the DMN is disturbed, progressing with duration of heroin use and correspond to decision making deficits |
| **Li et al., 2015 [63]** | FC: ICA | 26 | 0 | Study of 13 heroin relapsers and 13 abstainers; relapsers exhibit decreased FC in the left inferior temporal gyrus, right superior occipital gyrus within the DMN; increased FC in front precuneus, right middle cingulum |
| **Qiu et al., 2016 [8]** | Local: ALFF | 14 | 14 | Decreased ALFF values in the left post-central gyrus (associated with the SN) and left middle occipital cortex are associated with reduced severity of withdrawal symptoms |
| **Li et al., 2016 [64]** | FC: ICA | 24 | 20 | Abnormal FC within the anterior subnetwork of DMN associated with basal heroin craving |
| **Wang et al., 2016 [44]** | FC: SBA | 30 | 30 | Lower interhemispheric FC with the insula as the seed region; less deviation in interhemispheric FC associated with decreased risk of a positive heroin urine test |

**Table 7 (continued)**
Rs-fMRI Studies of OUD

| Study | Method | #OUD | #HC | Findings for OUD |
|---|---|---|---|---|
| **Chang et al., 2016 [5]** | Local: ReHo | 40 | 0 | Altered ReHo values in right caudate within the ECN correlated with craving and relapse after one year of methadone treatment |
| **Liu et al., 2020 [7]** | Local: ALFF | 21 | 21 | Increased ALFF in the left middle frontal gyrus of the ECN and decreased ALFF in the left postcentral gyrus, which is considered to have connections to the SN |
| **Woisard et al., 2021 [41]** | FC: ICA | 25 | 25 | No significant group differences are found for DMN, SN, or right ECN; preliminary results suggest left ECN connectivity may differ between OUD and HC |
| **Xue et al., 2022 [10]** | Local: ReHo | 51 | 42 | Increased ReHo values in the right orbitofrontal cortex and bilateral posterior central cortex within the ECN for 25 patients under methadone treatment compared to 26 patients after short-term abstinence |
| **Abdulaev et al., 2023 [42]** | FC: SBA | 12 | 16 | Decreased FC between the SN and ECN, decreased FC between SN and DMN, and increased FC between the DMN and ECN |
| **Chen et al., 2023 [43]** | FC: ICA | 37 | 57 | Increased FC between the medial mPFC and the left middle temporal gyrus of the DMN associated with reduced withdrawal symptoms; increased connectivity of the DMN's mPFC with the SN's anterior insular and middle frontal gyrus associated with increased withdrawal symptoms |
| **Dandurand et al., 2025 [45]** | FC: SBA | 11 | 0 | Greater FC of the DMN and SN is found after initiation of buprenorphine treatment compared to baseline scan obtained three days before |

The findings from this study have significant clinical implications for understanding and treating OUD. Identifying the DMN and SN as key functional networks associated with local neural alterations in OUD highlights the role of disrupted intrinsic neural activity in the disorder. Leveraging ML modeling of time-frequency features extracted from rs-fMRI BOLD signals provides a novel, detailed approach to distinguishing OUD from HC or individuals with other substance use disorders, offering potential biomarkers for early-stage therapeutics development. This ML-based framework could support such early-stage markers of potential treatments that target specific neural hubs through neuromodulation or pharmacological interventions. Future research should validate these findings in larger, more diverse populations to improve generalizability. Longitudinal studies examining time-frequency features of DMN, SN, and ECN activity throughout treatment could further clarify their roles as potential biomarkers for recovery. Combining this ML framework with other modalities, such as behavioral or genetic data, could optimize its diagnostic and therapeutic potential, ultimately enhancing outcomes for individuals with OUD.

## 5. Limitations

A potential confounding factor is the substance use history in the OUD group. Individuals with OUD are known to often use multiple substances, which may influence the findings. Due to lack of availability of substance use data, we were unable to perform subgroup analyses to assess the potential effects of other drugs. Although our study has a relatively small sample size, it is comparatively larger than other similar studies in the literature. Only two studies in Table 8 have a larger sample size ([43] and [10]). However, [43] focuses on FC analysis while the other [10] examines OUD patients undergoing treatment. Future studies with larger sample sizes, detailed substance use history, and balanced demographic data are necessary to validate these findings.

## 6. Conclusion

This study, for the first time in the literature, proposes an ML approach to identify differentiated functional brain networks for OUD using time-frequency analysis of rs-fMRI BOLD signals. Existing rs-fMRI studies of OUD have primarily focused on FC analyses between different areas of the brain. The few studies of OUD that have investigated local neural activity within specific functional networks use ALFF or ReHo as a static measure of neural activity. However, more recent studies have found meaningful variations in the resting-state signal over the duration of an rs-fMRI scan. Consequently, this study helps to understand the effect of OUD in functional brain areas using data-driven ML and time-frequency analysis of local neural activity within the DMN, SN, and ECN.

Future studies may validate and use these findings to generate new hypotheses for exploring the underlying neurobiology of OUD. While ML has proven to be a useful tool for a wide variety of clinical analyses, we have not seen other studies of OUD using ML for fMRI time-frequency analysis. Our proposed approach and findings demonstrate the feasibility and utility of ML modeling for rs-fMRI BOLD time-frequency analysis in OUD, which we hope will help facilitate further data-driven research aimed at understanding OUD-related changes in the brain. Further investigation will be required to understand the unique and complementary contributions of FC and data-driven ML for time-frequency analysis of local neural activity in the resting brain with respect to OUD. Future research can build on these findings to develop more accurate and effective diagnostic and treatment strategies for OUD and other related disease conditions.

## CRediT Authorship Contribution Statement

**Ahmed Temtam:** Data curation, Formal analysis, Investigation, Methodology, Software, Visualization, Writing – original draft, Writing – review & editing. **Megan A. Witherow:** Formal analysis, Investigation, Methodology, Software, Visualization, Writing – original draft, Writing – review & editing. **Liangsuo Ma:** Data Curation, Methodology, Validation, Writing – review & editing. **M. Shibly Sadique:** Software. **F. Gerard Moeller:** Funding Acquisition, Project Administration, Resources, Supervision, Validation. **Khan M. Iftekharuddin:** Conceptualization, Funding Acquisition, Project Administration, Resources, Supervision, Writing – review & editing.

**Declaration of Competing Interest**

The authors declare that they have no known competing financial interests or personal relationships that could have appeared to influence the work in this paper.

**Acknowledgements**

This research is supported by National Institutes of Health (NIH) - Clinical and Translational Science Awards (CTSA) Grant No. UM1TR004360.